\documentclass[superscriptaddress,secnumarabic,nobibnotes,aps,prd,showkeys,showpacs,
nofootinbib,twocolumn,10pt]{revtex4}
\usepackage{subfigure}
\usepackage{bm}
\usepackage{amsmath}
\usepackage{amsfonts}
\usepackage{amssymb}

\newcommand{\be}{\begin{equation}}
\newcommand{\ee}{\end{equation}}

\newcommand{\mincir}{\raise
-3.truept\hbox{\rlap{\hbox{$\sim$}}\raise4.truept\hbox{$<$}\ }}
\newcommand{\magcir}{\raise
-3.truept\hbox{\rlap{\hbox{$\sim$}}\raise4.truept\hbox{$>$}\ }}

\def\bea{\begin{eqnarray}}
\def\eea{\end{eqnarray}}

\def\be{\begin{equation}}
\def\ee{\end{equation}}

\PassOptionsToPackage{linktocpage}{hyperref}
\def\case#1/#2{\textstyle\frac{#1}{#2}}

\def\be{\begin{equation}}
\def\ee{\end{equation}}
\def\bea{\begin{eqnarray}}
\def\eea{\end{eqnarray}}

\begin{document}

\title{Cosmological solutions in Ho\v{r}ava-Lifshitz Scalar Field theory}
\author{Andronikos Paliathanasis}
\email{anpaliat@phys.uoa.gr}
\affiliation{Institute of Systems Science, Durban University of Technology, Durban 4000,
South Africa}
\author{Genly Leon}
\email{genly.leon@ucn.cl}
\affiliation{Departamento de Matem\'{a}ticas, Universidad Cat\'{o}lica del Norte, Avda.
Angamos 0610, Casilla 1280 Antofagasta, Chile}

\begin{abstract}
We perform a detailed study of the integrability of the Ho\v{r}ava-Lifshitz
scalar field cosmology in a Friedmann--Lema\^{\i}tre--Robertson--Walker
background spacetime. The approach we follow to determine the integrability
is that of Singularity Analysis. More specifically, we test if the
gravitational field equations possess the Painlev\'{e} property. For the
exponential potential of the scalar field we are able to perform an analytic
explicit integration of the field equations and write the solution in terms
of a Laurent expansion and more specifically write the solution in terms of
Right Painlev\'{e} series.
\end{abstract}

\maketitle
\date{\today }

\section{Introduction}

Modified theories of gravity have been a subject of special interest over
recent years because they provide a geometric approach to the description of
 observable phenomena. In modified theories of gravity new geometric
quantities are introduced into the Einstein-Hilbert Action of General
Relativity (GR). The new terms in the gravitational action provide new
components of geometric origin in the field equations which change the
dynamics such as the solutions of the latter to describe the phenomena
observed. In this work, we are interested specifically in the Ho\v{r}%
ava-Lifshitz (HL) gravity \cite{hor3}. HL gravity is a power-counting
renormalization theory with consistent ultra-violet behaviour exhibiting an
anisotropic Lifshitz scaling between time and space at the ultra-violet
limit. HL theory can provide Einstein's GR as a critical point.

The Einstein-Aether theory \cite{DJ,DJ2,Carru} is related with HL theory in
the classical limit \cite{esf}. In Einstein-Aether theory the kinematic
quantities of a time-like vector field (the Aether) are introduced into the
gravitational Action Integral.  The theory preserves locality and covariance
while it contains GR. Now when the aether field is hypersurface-orthogonal
then the classical limit of HL is recovered.  Hence every
hypersurface-orthogonal Einstein-\ae ther solution is a solution of HL
gravity \cite{jac01}.

Lorentz=violated theories have been applied in various models of
gravitational physics \cite{Barrow,in1,in2}. Applications of HL theory cover
a wide range of subjects, from compact stars to cosmological studies \cite%
{Wang:2017brl,Nilsson:2018knn,or1,or2}. In \cite{kir0,kir1} it was found
that \ HL cosmology provides an alternative to inflation and that the
universe can be singularity-free.  For reviews of HL cosmology we refer the
reader to \cite{kir2,kir3}. Very few closed=form solutions exist in HL cosmology
with or without any matter source \cite{hsol1,hsol2,hsol3,hsol4,hsol5}.
However, according to our knowledge there are no known analytical solutions
in HL cosmology when a minimally=coupled scalar field contributes to the total
evolution of the universe.

Recently, a detailed study on the dynamics of the HL scalar field cosmology
was performed in \cite{genlyhl}. More specifically the phase space of HL
cosmology was examined for a wide range of scalar field potentials by means
of the powerful method of $f$-devisors. Applications were presented for the
exponential, the power-law and other potentials. Singular power-law
solutions or de Sitter solutions were found to be described at the critical
points. However, these are peculiar solutions as they do not exist for any
initial conditions and they describe only an approximation of the generic
solution for arbitrary initial conditions.

{In general, physical theories provide a large number of free
parameters or boundary conditions to be involved so that numerical solutions, even if
practicable, give no real idea for the properties of the differential
equations, which are defined by the theory.  This means that analytic
techniques are essential for the study of real-world problems. Hence, the
knowledge for the existence and the determination of analytical or exact
solutions for a given dynamical system is important for the detailed
study and understanding of the dynamical system. More specifically, when a
dynamical system is integrable, then know that trajectories of (numerical)
solutions correspond to \textquotedblleft real\textquotedblright\ solutions
for the given dynamical system, or when it is feasible, we can write the
analytic solution by using closed-form functions.} There are many methods to
study the integrability of a dynamical system. In terms of Hamiltonian
system usually in physics we refer to the Liouville integrability; where the
Hamiltonian system admits sufficient number of conservation laws in order to
be solved by quadratures \cite{arnold1}. There is a plethora of cosmological
models which are Liouville integrable, for instance see \cite%
{ns01,ns02,ns03,ns04,ns05,ns06} and references therein. The importance of
the Liouville integrable cosmological models, apart that the analytical
solution can be determined, it is that it is possible that the conservation
laws to be applied to perform a canonical quantization which is a specific
approach of quantum cosmology \cite{q01,q02,q03,q04}.

Although, there are dynamical systems where they are not Liouville
integrable, but there is an analytical way to perform an explicit
integration of the dynamical system. One of the alternative methods is the
singularity analysis which is the main mathematical tool that we apply in
this work. The modern treatment of singularity analysis is summarized in the
Ablowitz, Ramani and Segur (ARS) algorithm \cite{Abl1,Abl2,Abl3} which
provide us with the information if a given differential equation passes the
Painlev\'{e} test and consequently if the solution of the differential
equation can be written as a Laurent expansion around a movable singularity.
Singularity analysis is a powerful method which has lead to the
determination of analytic solutions of various cosmological models in GR
\cite{miritzis,helmi,cots1,cots2,pcosm,bun1,bun2} or in modified theories of
gravity \cite{cots,sinFR,sinFR2,sinFT,sinS01,sinr01,sinr02}. The plan of the
paper is as follows.

In Section \ref{sec2}, the formal theory of HL cosmology under the
detailed-balance condition is presented. While the field equations of HL
scalar field cosmology are given. Moreover we derive the equivalent field
equations in the dimensionless variables by using the $H$-normalization
approach \cite{Leon:2009rc, genlyhl}. By using the $H$-normalization we
present the dynamical systems of our consideration which we study in terms
of integrability. In Section \ref{sec3}, we briefly discuss the ARS
algorithm. The main results of our analysis and the new analytic solutions
in HL cosmology are presented in Section \ref{sec4}. In Section \ref{sec5},
we discuss our results and draw our conclusions.

\section{Ho\v{r}ava-Lifshitz scalar field cosmology}

\label{sec2}

The gravitational Action Integral in Ho\v{r}ava-Lifshitz gravity under the
detailed-balance condition is given by the following expression \cite{kir1}
\begin{eqnarray}
S_{g} &=&\int dtd^{3}x\sqrt{g}N\left\{ \frac{2}{\kappa ^{2}}%
(K_{ij}K^{ij}-\lambda K^{2})\right.   \notag \\
&+&\left. \frac{\kappa ^{2}}{2w^{4}}C_{ij}C^{ij}-\frac{\kappa ^{2}\mu }{%
2w^{2}}\frac{\epsilon ^{ijk}}{\sqrt{g}}R_{il}\nabla _{j}R_{k}^{l}+\frac{%
\kappa ^{2}\mu ^{2}}{8}R_{ij}R^{ij}\right.   \notag \\
&-&\left. \frac{\kappa ^{2}\mu ^{2}}{8(3\lambda -1)}\left[ \frac{1-4\lambda
}{4}R^{2}+\Lambda R-3\Lambda ^{2}\right] \right\} ,  \label{acct}
\end{eqnarray}%
where the underlying geometry is written as
\begin{equation}
ds^{2}=-N^{2}dt^{2}+g_{ij}(dx^{i}+N^{i}dt)(dx^{j}+N^{j}dt),  \label{le.01}
\end{equation}%
in which the lapse and shift functions are respectively $N$ and $N_{i}$. The
spatial metric is given by $g_{ij}$, and roman letters indicate spatial
indices. Tensor $K_{ij}$ is the extrinsic curvature defined as $\ $%
\begin{equation}
K_{ij}=\left( {\dot{g_{ij}}}-\nabla _{i}N_{j}-\nabla _{j}N_{i}\right) /2N
\label{le.02}
\end{equation}%
and $C^{ij}$ denotes the Cotton tensor%
\begin{equation}
C^{ij}=\epsilon ^{ijk}\nabla _{k}\bigl(R_{i}^{j}-R\delta _{i}^{j}/4\bigr)/%
\sqrt{g}.  \label{le.03}
\end{equation}%
The covariant derivatives are defined with respect to the spatial metric $%
g_{ij}$. Finally, $\epsilon ^{ijk}$ is the totally antisymmetric unit
tensor, $\lambda $ is a dimensionless constant and the quantities $w$ and $%
\mu $ are constants. For simplicity we select to work with units where $%
\kappa ^{2}=8\pi G=1$.

According to the cosmological principle, the universe in large scales is
homogeneous and isotropic, consequently the line element (\ref{le.01})
reduces to the Friedmann--Lema\^{\i}tre--Robertson--Walker (FLRW) spacetime
where $N^{i}=0$ and $g_{ij}=a^{2}(t)\gamma _{ij}$ where now~$\gamma _{ij}$
denotes the maximally symmetic space of constant curvature $k$, i.e.
\begin{equation}
~\gamma _{ij}dx^{i}dx^{j}=\frac{dr^{2}}{1-kr^{2}}+r^{2}d\Omega _{2}^{2},
\label{le.04}
\end{equation}%
where $k=-1,0,+1$ and $d\Omega _{2}$ is the two-sphere.

In addition, we assume the contribution of a scalar field in the universe
with Action Integral
\begin{equation}
S_{\phi }=\int dtd^{3}x\sqrt{g}N\left[ \frac{3\lambda -1}{4}\frac{\dot{\phi}%
^{2}}{N^{2}}-V(\phi )\right] ,  \label{le.05}
\end{equation}%
where for the scalar field it has been considered that $\phi $ inherits the
symmetries of the FLRW spacetime.

Variation with respect to the metric in the Action Integrals provides the
second-order differential equations%
\begin{eqnarray}
3H^{2} &=&\frac{1}{2(3\lambda -1)}\left[ \frac{3\lambda -1}{4}\dot{\phi}%
^{2}+V(\phi )\right]   \notag \\
&&+\frac{3}{16(3\lambda -1)^{2}}\left[ -\frac{\mu ^{2}k^{2}}{a^{4}}-{\mu
^{2}\Lambda ^{2}}+\frac{2\mu ^{2}\Lambda k}{a^{2}}\right] ,  \label{le.06a1}
\end{eqnarray}%
\begin{align}
& 2\dot{H}+3H^{2}=-\frac{1}{2(3\lambda -1)}\left[ \frac{3\lambda -1}{4}\dot{%
\phi}^{2}-V(\phi )\right]   \notag \\
& -\frac{1}{16(3\lambda -1)^{2}}\left[ -\frac{\mu ^{2}k^{2}}{a^{4}}+{3\mu
^{2}\Lambda ^{2}}-\frac{2\mu ^{2}\Lambda k}{a^{2}}\right] ,  \label{le.07}
\end{align}%
where $H=\frac{\dot{a}}{a}$.

On the other hand, variation with respect to the scalar field provides the
Klein-Gordon equation
\begin{equation}
\ddot{\phi}+3H\dot{\phi}+\frac{2V^{\prime }(\phi )}{3\lambda -1}=0.
\label{le.08}
\end{equation}

The latter field equations can be written equivalently as follows
\begin{equation}
3H^{2}=\rho _{\phi }+\rho _{HL},  \label{le.09}
\end{equation}%
\begin{equation}
-2\dot{H}-3H^{2}=p_{\phi }+p_{HL},  \label{le.10}
\end{equation}%
and%
\begin{equation}
\dot{\rho}_{\phi }+3H\left( \rho _{\phi }+p_{\phi }\right) =0,  \label{le.11}
\end{equation}%
where we have defined
\begin{align}
& \rho _{\phi }=\frac{1}{2(3\lambda -1)}\left[ \frac{3\lambda -1}{4}\dot{\phi%
}^{2}+V(\phi )\right] ,  \notag \\
& p_{\phi }=\frac{1}{2(3\lambda -1)}\left[ \frac{3\lambda -1}{4}\dot{\phi}%
^{2}-V(\phi )\right] ,  \label{le.12}
\end{align}%
and%
\begin{equation}
\rho _{HL}=\frac{3}{16(3\lambda -1)^{2}}\left[ -\frac{\mu ^{2}k^{2}}{a^{4}}-{%
\mu ^{2}\Lambda ^{2}}+\frac{2\mu ^{2}\Lambda k}{a^{2}}\right] ,
\label{le.13}
\end{equation}%
\begin{equation}
p_{HL}=\frac{1}{16(3\lambda -1)^{2}}\left[ -\frac{\mu ^{2}k^{2}}{a^{4}}+{%
3\mu ^{2}\Lambda ^{2}}-\frac{2\mu ^{2}\Lambda k}{a^{2}}\right] .
\label{le.14}
\end{equation}

In this work we focus on the integrability of the dynamical system (\ref%
{le.06a1})-(\ref{le.07}). However, to follow a similar singularity analysis
as the one performed for the Einstein-Aether theory in \cite{pallatta} we
prefer to work with $H$-normalized variables.

$H$-normalized variables defined as ${x,y,z,u}$ in \eqref{le.15} were used
in \cite{Leon:2009rc} to analyze the dynamics of HL cosmology in presence of
an scalar field with exponential potential. That analysis can be extended
with the variable $s,$ and the function $f(s)$ to analyze potentials beyond
the exponential potential like in \cite{genlyhl} by introducing the
quantities $s$ and $f$ defined in \eqref{le.15}. Hence, the dimensionless
dynamical system \cite{genlyhl} defined by the new variables $\left\{
x,y,z,u,s\right\} $%
\begin{align}
& x=\frac{\dot{\phi}}{2\sqrt{6}H},\quad y=\frac{\sqrt{V(\phi )}}{\sqrt{6}H%
\sqrt{3\lambda -1}},  \notag \\
& z=\frac{\mu }{4(3\lambda -1)a^{2}H}~,\quad u=\frac{\Lambda \mu }{%
4(3\lambda -1)H},  \notag \\
& s=-\frac{V^{\prime }(\phi )}{V(\phi )},\quad f\left( s\right) \equiv \frac{%
V^{\prime \prime }(\phi )}{V(\phi )}-\frac{V^{\prime 2}(\phi )}{V(\phi )^{2}}%
,  \label{le.15}
\end{align}%
where function $f\left( s\right) $ is defined by the specific form of the
potential $V\left( \phi \right) .$ As a new independent variable it is
assumed the number of e-fold $\tau =\ln \left( \frac{a}{a_{0}}\right) $
allows to recast the cosmological equations for arbitrary potentials in a
closed form. We remark, that for the exponential potential $V\left( \phi
\right) =V_{0}e^{\sigma \phi }$, function $f\left( s\right) $ vanishes and $%
s=\sigma $, while for the power-law potential $V\left( \phi \right)
=V_{0}\phi ^{2n}$ \ function $f\left( s\right) $ is defined as $f(s)=-\frac{%
s^{2}}{2n}$.

At this point it is important to mention that while the analysis of critical
points in \cite{genlyhl} for the HL scalar field cosmology is for arbitrary
potential, in order to study the integrability of the field equations by
applying the ARS algorithm the potential function has to be specified. The
exponential and power-law potentials are two well-known and well-studied
potentials in GR while they can approximate the functional behaviour of
other nonlinear potentials in their limits; for instance, the hyperbolic
potential $V\left( \phi \right) =V_{0}\sinh \left( \sigma \phi \right) ^{2n}$
where for small values of $\phi $ is approximated by the power law potential
$V\left( \phi \right) \simeq \phi ^{2n}$ and for large values of $\phi $ the
hyperbolic potential has an exponential behaviour.

\section{Singularity analysis}

\label{sec3}

The integration of a differential equation is performed by the global
knowledge of the general solution independently from the local solutions
provided by Cauchy's existence theorem. A differential equation which passes
the Painlev\'{e} test we shall say that it is integrable. However, because
not all the integrable differential equations have the Painlev\'{e}
property, the latter property when it is exists it is better to be referred
as the uniformizability of the general equation for the given differential
equation \cite{Rconte}.

In order to study if a given differential equation admits the Painlev\'{e}
property we apply the ARS algorithm which is summarized in the three main
steps: (a) determine the leading-order term which describe a movable
singularity; (b) determine the resonances which gives the position of the
integration constants and (c) write a Painlev\'{e} Series with exponent and
step as given in steps (a), (b) and test if solves the differential
equation, the latter it is called the consistency test.

There are various criteria and conditions which should be satisfied in the
ARS algorithm in order a differential equation to have the Painlev\'{e}
property, for instance, the number of the resonances should be equal with
the number of the degrees of freedom of the differential equation, while one
of the resonances should be the $r=-1$, otherwise the singularity determined
at step (a) is not movable. Moreover, if the resonances are positive, the
Painlev\'{e} Series is written by a Right Laurent expansion, when the
resonances are negative the Painlev\'{e} Series is given by a Left Laurent
expansion, otherwise the Painlev\'{e} Series is given by a mixed Laurent
expansion. \ For more details on the criteria of the ARS algorithm and the
interpretation of the algorithm in the complex plane we refer the reader to
\cite{Rconte,buntis}.

\section{Explicit integration in Ho\v{r}ava-Lifshitz Cosmology}

\label{sec4}

In this section we study if the field equations of HL scalar field cosmology
(\ref{le.06a1})-(\ref{le.08}) possess the Painlev\'{e} property, when the
latter is true we perform an explicit integration around the movable
singularity and we present the generic solution in terms of Laurent
expansion. We divide our analysis in the four different cases as studied in
\cite{genlyhl} for arbitrary potentials, and previously in \cite{Leon:2009rc}
for exponential potential.

In case A we assume that the underlying geometry is spatially flat $\left(
k=0\right) $ and there is no cosmological constant term $\left( \Lambda
=0\right) $, case B is with $k\neq 0$ and $\Lambda =0$. Cases C and D are
with $\Lambda \neq 0$ while for the spatial curvature it holds $k=0$ and $%
k\neq 0$ respectively. Last but not least for the scalar field potential we
consider the exponential potential in where $f\left( s\right) \equiv 0$ and
the power law potential with $f(s)=-\frac{s^{2}}{2n}$.

\subsection{Case A: $k=0,~\Lambda =0$}

For the first case of our analysis where the spatial curvature of the FLRW
spacetime vanishes and there is not a cosmological constant term, the
gravitational field equations in the dimensionless variables (\ref{le.15})
form the following three dimensional first-order differential equations \cite%
{genlyhl}

\begin{align}
& \frac{dx}{d\tau }=\left( 3x-\sqrt{6}s\right) \left( x^{2}-1\right) ,
\label{c1a} \\
& \frac{dz}{d\tau }=\left( 3x^{2}-2\right) z,  \label{c1b} \\
& \frac{ds}{d\tau }=-2\sqrt{6}xf(s),  \label{c1c}
\end{align}%
defined on the phase space $\{(x,z,s)\in \mathbb{R}^{3}:-1\leq x\leq 1\}.$

We continue our analysis by presenting the application of the ARS algorithm
for the dynamical system.

\subsubsection{Exponential potential}

For the exponential potential because the rhs of equation (\ref{c1c}) is
identically zero the dynamical system is reduced to a two-dimensional
first-order differential equations where $s\left( \tau \right) =const$.

We observe that the dynamical system (\ref{c1a}), (\ref{c1b}) for $s=const$.
can be easily integrated by quadratures. Indeed, from\ equation (\ref{c1a})
it follows that
\begin{align}
&\ln \left( \left( x\left( \tau \right) +1\right) ^{\frac{1}{6+2\sqrt{6}s}%
}\left( x\left( \tau \right) +1\right) ^{\frac{1}{6-2\sqrt{6}s}}\right) +
\notag \\
&\frac{3}{9-6s^{2}}\ln \left( \sqrt{6}s-3x\left( \tau \right) \right) =\tau
-\tau _{0}~,~s\neq \pm \frac{\sqrt{6}}{2},  \label{c1.01}
\end{align}%
or%
\begin{equation}
\frac{1}{12}\ln \left( \frac{x+1}{x-1}\right) -\frac{1}{6\left( x+1\right) }%
=\tau -\tau _{0}~,~s=-\frac{\sqrt{6}}{2},
\end{equation}%
or%
\begin{equation}
\frac{1}{12}\ln \left( \frac{x-1}{x+1}\right) +\frac{1}{6\left( x+1\right) }%
=\tau -\tau _{0}~,~s=-\frac{\sqrt{6}}{2},
\end{equation}%
while equation (\ref{c1b}) \ gives%
\begin{equation}
\ln z\left( \tau \right) =3\int x\left( t\right) ^{2}d\tau -2\left( \tau
-\tau _{1}\right) \text{.}  \label{c1.02}
\end{equation}

Notice that this system was already integrated at \cite{Leon:2009rc}, in a
rather different way:
\begin{eqnarray}
&&z(x)=z_{0}\left( \frac{3x-\sqrt{6}s}{3x_{0}-\sqrt{6}s}\right) ^{\frac{%
2(s^{2}-1)}{2s^{2}-3}}\left( \frac{x^{2}-1}{x_{0}^{2}-1}\right) ^{\frac{1}{%
6-4s^{2}}}\cdot   \notag \\
&&\cdot \exp \left\{ \frac{\sqrt{6}s\left[ \tanh ^{-1}(x)-\tanh ^{-1}(x_{0})%
\right] }{6s^{2}-9}\right\} , \\
&&\tau -\tau _{0}=\ln \left( \frac{\sqrt{6}-3x_{0}}{\sqrt{6}-3x}\sqrt{\frac{%
1-x^{2}}{1-x_{0}^{2}}}\right)   \notag \\
&&-\sqrt{\frac{2}{3}}\left( \tanh ^{-1}(x)-\tanh ^{-1}(x_{0})\right) .
\end{eqnarray}%
for $x_{0}-\sqrt{6}s\neq 0,x_{0}^{2}\neq 1,s\neq \pm \frac{\sqrt{6}}{2}$.

Let us now apply the ARS algorithm in the system (\ref{c1a}), (\ref{c1b}).
For the first step of the algorithm we substitute%
\begin{equation}
x\left( \tau \right) =x_{0}\tau ^{p}~,~z\left( \tau \right) =z_{0}\tau ^{q}
\label{c1.03}
\end{equation}%
and we find
\begin{equation}
px_{0}\tau ^{-1+p}-3x_{0}^{3}\tau ^{3p}+3x_{0}\tau ^{p}+\sqrt{6}%
sx_{0}^{2}\tau ^{2p}-\sqrt{6}s=0,  \label{c1.04}
\end{equation}%
\begin{equation}
\left( q\tau ^{-1+q}+2t^{q}-3x_{0}\tau ^{2p+q}\right) z_{0}=0.  \label{c1.05}
\end{equation}

In order to find the leading-order term we should balance at least two of
the exponents of powers of $\tau $. Note that in order a singularity to be
occurred parameters $p,q$ should be negative\footnote{%
This is not absolute, in the modern treatment the exponent $p$ of the
leading-order term can be also be a positive fractional number.}, while in
general $\tau \rightarrow \tau -\tau _{0}$ where~$\tau _{0}$ denotes the
position of the singularity where without loss of generality we consider $%
\tau _{0}=0$.

Hence, from (\ref{c1.05}) it follows that $-1+q=2p+q$ or $q=2p+q$. The
second case provide $p=0$ which we reject, while from the other case we
derive $p=-\frac{1}{2}$ where the leading order terms have exponents $\tau
^{-\frac{3}{2}}$ which they cancel when $x_{0}^{2}=\sqrt{\frac{q}{3}}$.
Moreover from expression (\ref{c1.04}) we find that the leading-order terms
are that with power $\tau ^{-\frac{3}{2}}$ and coefficient $-\frac{x_{0}}{2}%
\left( 1+6x_{0}^{2}\right) $ where by demanding the latter to be zero we
find that $q=0$ or $q=-\frac{1}{2}$. We conclude, that the leading order
behaviour is%
\begin{equation}
x\left( \tau \right) =\pm \frac{i}{\sqrt{6}}\tau ^{-\frac{1}{2}}~,~z\left(
\tau \right) =z_{0}\tau ^{-\frac{1}{2}}\text{.}  \label{c1.06}
\end{equation}

In order to find the resonances $r$ we substitute
\begin{equation}
x\left( \tau \right) =\pm \frac{i}{\sqrt{6}}\tau ^{-\frac{1}{2}}+m\tau ^{-%
\frac{1}{2}+r}~,~~z\left( \tau \right) =z_{0}\tau ^{-\frac{1}{2}}+\nu \tau
^{-\frac{1}{2}+r}  \label{c1.07}
\end{equation}%
in (\ref{c1a}), (\ref{c1b}) from where we get
\begin{align}
&0 =-\frac{x_{0}}{2}\left( 1+6x_{0}^{2}\right) \tau ^{-\frac{3}{2}}+\sqrt{6}%
sx_{0}^{2}\tau ^{-1}+3x_{0}\tau ^{-\frac{1}{2}}-\sqrt{6}s+  \notag \\
&+\left( \frac{1}{2}\left( 2r-1-18x_{0}^{2}\right) t^{-\frac{3}{2}+r}+2\sqrt{%
6}sx_{0}\tau ^{-1+r}+3\tau ^{-\frac{1}{2}+r}\right) m  \notag \\
& +O\left( m^{2}\right),  \label{c1.08}
\end{align}%
\begin{align}
&0 =\left( -\frac{1+6x_{0}^{2}}{2}\tau ^{-\frac{3}{2}}+2t^{-\frac{1}{2}%
}\right) +\left( -6x_{0}\tau ^{-\frac{3}{2}+r}\right) m+  \notag \\
&+\left( \left( r-\frac{1}{2}-3x_{0}^{2}\right) \tau ^{-\frac{3}{2}+r}+2\tau
^{-\frac{1}{2}+r}\right) n  \notag \\
& +O\left( m^{2},n^{2},mn\right) ,  \label{c1.09}
\end{align}%
where $x_{0}=\pm \frac{i}{\sqrt{6}}$.

From the latter system we define the two-dimensional matrix%
\begin{equation}
A=%
\begin{pmatrix}
\frac{1}{2}\left( 2r-1-18x_{0}^{2}\right) & 0 \\
\left( -6x_{0}\tau ^{-\frac{3}{2}+r}\right) & \left( r-\frac{1}{2}%
-3x_{0}^{2}\right)%
\end{pmatrix}%
,  \label{c1.10}
\end{equation}%
whose requirement $\det A=0$ says that parameters $m$ and $n$ are arbitrary.
We find $\det A=r\left( r+1\right) $ from where we calculate the two
resonances to be
\begin{equation}
r=-1\text{ and }r=0.  \label{c1.11}
\end{equation}%
The existence or $r=-1$ indicate that the singularity is movable while $r=0$
says that the second integration constant is the coefficient parameter $%
z_{0}~$\ in (\ref{c1.06}). Since the two integration constants have been
determined we can say that the system (\ref{c1a}), (\ref{c1b}) possess the
Painlev\'{e} property. The algebraic solution is expressed by the\ Laurent
expansions%
\begin{align}
& x\left( \tau \right) =\pm \frac{i}{\sqrt{6}}\tau ^{-\frac{1}{2}%
}+\sum_{j=1}x_{j}\tau ^{-\frac{1}{2}+j},  \notag \\
& z\left( \tau \right) =z_{0}\tau ^{-\frac{1}{2}}+\sum_{j=1}z_{j}\tau ^{-%
\frac{1}{2}+j},  \label{c1.12}
\end{align}%
where the first coefficient terms for the $x_{0}=+\frac{i}{\sqrt{6}}$
coefficient are determined to be
\begin{equation}
x_{1}=\frac{1}{3}\sqrt{\frac{2}{3}}s~,~x_{2}=-\frac{i}{6\sqrt{6}}\left(
9+2s^{2}\right),~...  \label{c1.13}
\end{equation}%
\begin{equation}
z_{1}=\frac{4}{3}sz_{0}i~,~z_{2}=-\frac{z_{0}}{6}\left( 3+2s^{2}\right),~...
\label{c1.14}
\end{equation}

Therefore, the Laurent expansions (\ref{c1.12}) pass the third-step of the
ARS algorithm and we conclude that the two-dimensional dynamical system (\ref%
{c1a}), (\ref{c1b}) possess the Painlev\'{e} property. In the following we
avoid the presentation of the calculations and we give directly the main
results which follow from our analysis.

\subsubsection{Power-law potential}

We continue with the study of the three-dimensional dynamical system (\ref%
{c1a})-(\ref{c1c}) where $f(s)=-\frac{s^{2}}{2n}$. The leading-order terms
are found to be%
\begin{equation}
x\left( \tau \right) =x_{0}\tau ^{-\frac{1}{2}}~,~z\left( \tau \right)
=z_{0}\tau ^{-\frac{1+n}{2}}~,~s\left( \tau \right) =s_{0}\tau ^{-\frac{1}{2}%
}  \label{c1.15}
\end{equation}%
in which%
\begin{equation}
x_{0}^{2}=-\frac{1+n}{6}~,~s_{0}=-\frac{n}{2\sqrt{6}x_{0}}~\text{and~}z_{0}=%
\text{arbitrary.}  \label{c1.16}
\end{equation}%
where $n\neq 0,-1,-3$. \ The resonances are the zeros of the polynomial
equation $\left( 1+2r\right) \left( r+1\right) r=0$ which are%
\begin{equation}
r=-1~,~r=0\text{ and }r=-\frac{1}{2}\text{. }  \label{c1.17}
\end{equation}

At this case the generic solution is given by the following mixed Laurent
expansions%
\begin{equation}
x\left( \tau \right) =\sum_{j=1}\bar{x}_{j}\tau ^{-\frac{1}{2}-j}+x_{0}\tau
^{-\frac{1}{2}}+\sum_{j=1}x_{j}\tau ^{-\frac{1}{2}+j}  \label{c1.18}
\end{equation}%
\begin{equation}
z\left( \tau \right) =\sum_{j=1}\bar{z}_{j}\tau ^{-\frac{1+n}{2}%
-j}+z_{0}\tau ^{-\frac{1+n}{2}}+\sum_{j=1}z_{j}\tau ^{-\frac{1+n}{2}+j}
\label{c1.19}
\end{equation}%
\begin{equation}
s\left( \tau \right) =\sum_{j=1}\bar{s}_{j}\tau ^{-\frac{1}{2}-j}+s_{0}\tau
^{-\frac{1}{2}}+\sum_{j=1}s_{j}\tau ^{-\frac{1}{2}+j}  \label{c1.20}
\end{equation}%
where $x_{0},~z_{0}$ and $s_{0}$ given in (\ref{c1.18}).

In order to perform the consistency test we select $n=1$, and for $x_{0}=%
\frac{i}{\sqrt{3}}~$we find that $x_{j}=0~,~s_{j}=0$ for every value of $j$
and the third integration constant is parameter $\bar{s}_{1}$, recall that
the other two integration constants are the position of the movable
singularity $\tau _{0}$ and the coefficient $z_{0}$. For negative values of $%
n$ the consistency test fails. Hence the system posses the Painlev\'{e}
property only for $n>0$.

We proceed with the second case of our analysis.

\subsection{Case B: $k\neq 0,~\Lambda =0$}

In the presence of curvature the dimensionless field equations form the
following three-dimensional dynamical system \cite{genlyhl}
\begin{align}
& \frac{dx}{d\tau }=x\left( 3x^{2}-2z^{2}-3\right) +\sqrt{6}s\left(
1-x^{2}+z^{2}\right) ,  \label{c2.01} \\
& \frac{dz}{d\tau }=z\left[ 3x^{2}-2\left( z^{2}+1\right) \right] ,
\label{c2.02} \\
& \frac{ds}{d\tau }=-2\sqrt{6}xf(s),  \label{c2.03}
\end{align}%
which is defined on the phase space $\{(x,z,s)\in \mathbb{R}%
^{3}:x^{2}-z^{2}\leq 1\}.$

\subsubsection{Exponential potential}

For the exponential potential and for the two-dimensional system (\ref{c2.01}%
), (\ref{c2.02}) the leading-order behaviour is found to be%
\begin{equation}
x\left( \tau \right) =\pm \sqrt{\frac{4z_{0}-1}{6}}\tau ^{-\frac{1}{2}%
}~,~z\left( \tau \right) =z_{0}\tau ^{-\frac{1}{2}}  \label{c2.04}
\end{equation}%
in which $z_{0}$ is an arbitrary constant. The resonances are determined to
be again $r=-1$ and $r=0$ as in the spatially flat case. The solution is
expressed in the right Laurent expansions
\begin{align}
& x\left( \tau \right) =\pm \frac{\sqrt{4z_{0}-1}}{\sqrt{6}}\tau ^{-\frac{1}{%
2}}+\sum_{j=1}x_{j}\tau ^{-\frac{1}{2}+j},  \notag \\
& z\left( \tau \right) =z_{0}\tau ^{-\frac{1}{2}}+\sum_{j=1}z_{j}\tau ^{-%
\frac{1}{2}+j},  \label{c2.05}
\end{align}%
where now the first coefficients are defined as%
\begin{align}
& x_{1}=\frac{1}{3}\sqrt{\frac{2}{3}}s\left( 1+2z_{0}^{2}\right) \left(
1+8z_{0}^{2}\right) ,  \notag \\
& z_{1}=\frac{4}{3}sz_{0}\left( 1+2z_{0}^{2}\right) \sqrt{4z_{0}^{2}-1}~,...
\label{c2.06}
\end{align}

We continue with the three-dimensional system defined by the power-law
potential.

\subsubsection{Power-law potential}

For the power-law potential the first step of the ARS algorithm for the
dynamical system (\ref{c2.01})-(\ref{c2.03}) provides the leading-order
behaviour%
\begin{align}
& x\left( \tau \right) =\pm \sqrt{\frac{4z_{0}^{2}-1}{6}}\tau ^{-\frac{1}{2}%
}~,~z\left( \tau \right) =z_{0}\tau ^{-\frac{1}{2}},  \notag \\
& s\left( \tau \right) =-\frac{\left( n+1\right) p}{\sqrt{6}x_{0}}\tau ^{-%
\frac{1}{2}}~,~z_{0}=\pm \frac{i}{\sqrt{2}},  \label{c2.07}
\end{align}%
for $n\neq 0,-1$.

As far as the resonances are concerned they are derived%
\begin{equation}
r=-1~,~r=-\sqrt{6}~,~r=\frac{1}{2}+\frac{\sqrt{6}}{n}  \label{c2.08}
\end{equation}%
where we conclude that the given dynamical system does not possess the
Painlev\'{e} property.

\subsection{Case C: $k=0,~\Lambda \neq 0$}

For the spatially flat background space and in the presence of the
cosmological constant $\Lambda $, the field equations reduce to the
following dimensionless system \cite{genlyhl}
\begin{align}
& \frac{dx}{d\tau }=\sqrt{6}s\left( u^{2}-x^{2}+1\right) +3x\left(
x^{2}-1\right) ,  \label{c3.01} \\
& \frac{du}{d\tau }=3ux^{2},  \label{c3.02} \\
& \frac{ds}{d\tau }=-2\sqrt{6}xf(s).  \label{c3.03}
\end{align}%
defined on the phase space $\{(x,u,s)\in \mathbb{R}^{3}:x^{2}-u^{2}\leq 1\}.$

\subsubsection{Exponential potential}

The two-dimensional dynamical system (\ref{c3.01}), (\ref{c3.02}) with $%
s=const$. passes the Painlev\'{e} test. The solution is given by the right
Laurent expansions%
\begin{align}
& x\left( \tau \right) =\pm \frac{i}{\sqrt{6}}\tau ^{-\frac{1}{2}%
}+\sum_{j=1}x_{j}\tau ^{-\frac{1}{2}+j}~,  \notag \\
& u\left( \tau \right) =u_{0}\tau ^{-\frac{1}{2}}+\sum_{j=1}u_{j}\tau ^{-%
\frac{1}{2}+j}  \label{c3.04}
\end{align}%
where the constants of integrations are $u_{0}$ and the position of the
movable singularity $\tau _{0}$.

By replacing (\ref{c3.04}) in the dynamical system the first coefficients
are calculated (for $x_{0}=\frac{i}{\sqrt{6}}$ )
\begin{align}
&x_{1}= \frac{\sqrt{6}s\left( 1+6u_{0}^{2}\right) }{9}~,  \notag \\
& x_{2}=\frac{i}{6\sqrt{6}}\left( s^{2}\left(
360u_{0}^{4}+48u_{0}^{2}-2\right) -9\right) ~,~... \\
& u_{1}= i\frac{4}{3}su_{0}\left( 1+6u_{0}^{2}\right)~,  \notag \\
& u_{2}=-\frac{u_{0}}{6}\left( s^{2}\left( 504u_{0}^{4}+96u_{0}^{2}+2\right)
-9\right) ~,~...
\end{align}

\subsubsection{Power-law potential}

In the case of the power-law potential and for the three-dynamical system (%
\ref{c3.01})-(\ref{c3.03}) with $n\neq -1,0$; from the first step of the ARS
algorithm we determine the leading order behaviour%
\begin{equation}
x\left( \tau \right) =\pm \frac{i}{\sqrt{6}}\tau ^{-\frac{1}{2}}~,~u\left(
\tau \right) =\pm \frac{i}{\sqrt{6}}\tau ^{-\frac{1}{2}}~,~s\left( \tau
\right) =\frac{i}{2}\tau ^{-\frac{1}{2}}.
\end{equation}%
The resonances are found to be the solutions of the polynomial equation $%
\left( r+1\right) \left( 2r+1\right) \left( r-n\right) =0$, which gives%
\begin{equation*}
r=-1~,~r=-\frac{1}{2}~,~r=n.
\end{equation*}

In order to write the Laurent expansions and perform the consistency test
the power $n$ should be defined. We have studied various values of $n$,
positives and negatives, integers and fractional, and we found that in the
possible cases the Laurent expansions fail at the consistency test. \ Hence,
we conclude that the dynamical system (\ref{c3.01})-(\ref{c3.03}) does not
pass the Painlev\'{e} test.

\subsection{Case D: $k\neq 0,~\Lambda \neq 0$}

The dimensionless dynamical system which describe the field equations in the
last case of our consideration consists by the following four first-order
differential equations \cite{genlyhl}
\begin{align}
\frac{dx}{d\tau }& =\sqrt{6}s\left[ -x^{2}+(u-z)^{2}+1\right]  \notag \\
& +x\left[ 3x^{2}+2(u-z)z-3\right] ,  \label{c4.01} \\
\frac{dz}{d\tau }& =z\left[ 3x^{2}+2(u-z)z-2\right] ,  \label{c4.02} \\
\frac{du}{d\tau }& =u\left[ 3x^{2}+2(u-z)z\right] ,  \label{c4.03} \\
\frac{ds}{d\tau }& =-2\sqrt{6}xf(s),  \label{c4.04}
\end{align}%
defined on the phase space $\{(x,z,u,s)\in \mathbb{R}^{3}:x^{2}-(u-kz)^{2}%
\leq 1\}.$

\subsubsection{Exponential potential}

For the three-dimensional dynamical system (\ref{c4.01})-(\ref{c4.03}) where
parameter $s$ is constant, we find the leading-order behaviour%
\begin{equation}
x\left( \tau \right) =x_{0}\tau ^{-\frac{1}{2}}~,~z\left( \tau \right)
=z_{0}\tau ^{-\frac{1}{2}}~,~u\left( \tau \right) =u_{0}\tau ^{-\frac{1}{2}},
\label{c4.05}
\end{equation}%
where $x_{0},~z_{0}$ are arbitrary constants and $u_{0}=\frac{%
4z_{0}^{2}-1-6x_{0}^{2}}{4z_{0}}$, with $4z_{0}^{2}-1-6x_{0}^{2}\neq 0$.
Because there are already three arbitrary constants, including the position
of the singularities someone will expect to find two resonances with value
zero.

Indeed, by replacing
\begin{align}
& x\left( \tau \right) =x_{0}\tau ^{-\frac{1}{2}}+m\tau ^{-\frac{1}{2}%
+r}~,~z\left( \tau \right) =z_{0}\tau ^{-\frac{1}{2}}+\nu \tau ^{-\frac{1}{2}%
+r}~,  \notag \\
& u\left( \tau \right) =u_{0}\tau ^{-\frac{1}{2}}+\kappa \tau ^{-\frac{1}{2}%
+r},  \label{c4.06}
\end{align}%
in the dynamical system (\ref{c4.01})-(\ref{c4.03}) we derive the resonances
are the zeros of the polynomial equation $r^{2}\left( r+1\right) $. We
conclude that the dynamical system (\ref{c4.01})-(\ref{c4.03}) with $s$
constant $\ $possess the Painlev\'{e} property. The algebraic solution is
given by the following Laurent expansions%
\begin{align}
& x\left( \tau \right) = x_{0}\tau ^{-\frac{1}{2}}+\sum_{j=1}x_{j}\tau ^{-%
\frac{1}{2}+j}~,  \notag \\
& z\left( \tau \right) = z_{0}\tau ^{-\frac{1}{2}}+\sum_{j=1}z_{j}\tau ^{-%
\frac{1}{2}+j},  \label{c4.07} \\
& u\left( \tau \right) =\frac{4z_{0}^{2}-1-6x_{0}^{2}}{4z_{0}}\tau ^{-\frac{1%
}{2}}+\sum_{j=1}u_{j}\tau ^{-\frac{1}{2}+j}.  \label{c4.08}
\end{align}

\subsubsection{Power-law potential}

For the power-law potential the leading-order behaviour are found to be
\begin{align}
& x\left( \tau \right) =x_{0}\tau ^{-\frac{1}{2}}~,~z\left( \tau \right)
=z_{0}\tau ^{-\frac{1}{2}}~,  \notag \\
& u\left( \tau \right) =u_{0}\tau ^{-\frac{1}{2}}~,~s\left( \tau \right)
=s_{0}\tau ^{-\frac{1}{2}},
\end{align}%
with%
\begin{equation*}
z_{0}=-\frac{6x_{0}^{2}+1}{4x_{0}}~,~u_{0}=-\frac{2x_{0}^{2}+1}{4x_{0}}%
~,~s_{0}=-\frac{n}{2\sqrt{6}x_{0}},
\end{equation*}%
and $x_{0}$ arbitrary. The second step of the ARS algorithm provide the\
four resonances%
\begin{equation}
r=-1~,~r=0,~r=-\frac{1}{2}\text{ and }r=n.
\end{equation}

In a similar way with case C we have to define power-index $n$ in order to
perform the consistency test. We have performed the consistency test for
various rational numbers of $n$ and we found can conclude that the
four-dimensional dynamical system (\ref{c4.01})-(\ref{c4.04}) does not
posses the Painlev\'{e} property.

\section{Conclusion}

In this work we studied the integrability of the HL scalar field cosmology
in a FLRW background spacetime for the exponential and power-law potentials.
We performed our study for the dimensionless dynamical system under the $H$%
-normalization which is usually applied in the fixed point analysis of
gravitational dynamical systems. We categorized our study in for cases of
study according to the existence of the cosmological constant term $\Lambda $
and if the spatially curvature $k~$vanishes or not. More specifically the
four cases of study are: Case A: $k=0,~\Lambda =0,~$Case B: $k\neq
0,~\Lambda =0$, Case C:~$k=0,~\Lambda \neq 0$ and Case D: $k\neq 0,~\Lambda
\neq 0$.

The main mathematical tool that we applied for the study of the
integrability of the dimensionless field equations for the abovementioned
cases of study is that of the singularity analysis. In particular we
examined if the given gravitational dynamical system possess the Painlev\'{e}
property which tell us that an explicit analytic integration can be
performed where the solution is expressed in Laurent expansion.

As far as the power-law potential is concerned we found that only Case A
provides an integrable system. In contrary to the exponential potential
where the field equations always possesses the Painlev\'{e} property. Our
results are summarized in the following proposition:\textit{\
\textquotedblleft The gravitational field equations for the HL scalar field
cosmology in a FLRW background (\ref{le.06a1})-(\ref{le.08}) expressed in
the dimensionless variables in the }$H-$\textit{normalization (\ref{le.15})
pass always the Painlev\'{e} test when }$V\left( \phi \right)
=V_{0}e^{-\sigma \phi }$\textit{, and the equations can be explicitly
integrated by Laurent expansions, where in all cases the resonances are }$%
r=-1$\textit{\ and }$r=0$\textit{, where the rank of }$r=0$\textit{\ is
greater of equal to one\textquotedblright . }At this point we want to
mention that by applying the same analysis for the field equations beyond
the detailed-balance condition \ \cite{genlyhl} we found that the field
equations do not posses the Painlev\'{e} property.

The results of this analysis complement the dynamical study of HL scalar
field equations \cite{genlyhl} and the analysis in \cite{Leon:2009rc}.\ Last
but not least, this work contributes to the subject of integrability of
gravitational field equations in cosmological studies.

\label{sec5}

\begin{acknowledgments}
The authors want to thank the anonymous referee for valuable comments and suggestions which helped to improve the presentation of this work.
This work was supported by Comisi\'{o}n Nacional de Investigaci\'{o}n Cient%
\'{\i}fica y Tecnol\'{o}gica (CONICYT) through FONDECYT Iniciaci\'{o}n
11180126. GL thanks to Departmento de Matem\'{a}tica and to Vicerrector\'{\i}%
a de Investigaci\'{o}n y Desarrollo Tecnol\'{o}gico at Universidad Cat\'{o}%
lica del Norte for financial support.
\end{acknowledgments}

\end{document}